\documentclass[12pt,preprint]{aastex}

\shorttitle{Building Black Holes \& Bulges}
\shortauthors{T. Heckman, G. Kauffmann et al.}

\begin{document}


\title{Present-Day Growth of Black Holes and Bulges:\\the SDSS Perspective}


\author{Timothy M. Heckman\altaffilmark{1}, Guinevere Kauffmann\altaffilmark{2},Jarle Brinchmann \altaffilmark{2,3},
St\'ephane Charlot\altaffilmark{2,4}, Christy Tremonti\altaffilmark{5} and Simon D.M. White\altaffilmark{2}}
\email{heckman@pha.jhu.edu, gamk@mpa-garching.mpg.de}

\altaffiltext{1}{Center for Astrophysical Sciences, Department of Physics \& Astronomy,
Johns Hopkins University, Baltimore, MD 21218}
\altaffiltext{2} {Max Planck Institut fur Astrophysik, D-85748 Garching, Germany}
\altaffiltext{3} {Centro de Astrofisica da Universidade do Porto,
Rua das Estrelas, 4150-762 Porto, Portugal}
\altaffiltext{4}{Institut d'Astrophysique du CNRS, 98 bis Boulevard Arago, F-75014 Paris, France} 
\altaffiltext{5}{ Steward Observatory, 933 North Cherry Ave, Tuscon, AZ 85721}


\begin{abstract}

We investigate the accretion-driven growth of supermassive black holes
in the low-redshift Universe using 23,000 narrow-emission-line (``Type
2'') active galactic nuclei (AGN) and the complete sample of
123,000 galaxies in the Sloan Digital Sky Survey from which they were
drawn.  We use the stellar velocity dispersions
of the early type galaxies and AGN hosts to estimate their black
hole masses and we use the AGN [OIII]$\lambda$5007 emission line
luminosities to estimate black hole accretion rates.  We find that
most present-day accretion occurs onto black holes with masses less
than $10^8 M_{\odot}$ that reside in moderately massive galaxies
($M_* \sim 10^{10}$ to $10^{11.5} M_{\odot}$) with high stellar
surface mass densities ($\mu_* \sim 10^{8.5}$ to $10^{9.5} M_{\odot}$
kpc$^{-2}$) and young stellar populations.  The volume-averaged
accretion rates of low mass black holes ($< 3 \times 10^7 M_{\odot}$)
imply that this population is growing on a timescale that is comparable to the
age of the Universe.  Around half this growth takes place in AGN that
are radiating within a factor of five of the Eddington
luminosity. Such systems are rare, making up only 0.2 \% of the low
mass black hole population at the present day. The rest of the growth
occurs in lower luminosity AGN.  The growth timescale is more than 
two orders of magnitude longer for the population of the
most massive black holes in our
sample.  The volume-averaged ratio of star formation to black hole
accretion in bulge-dominated galaxies is $\sim 1000$, in remarkable
agreement with the observed ratio of stellar mass to black hole mass
in nearby galaxy bulges.  We conclude: a) that bulge formation and
black hole formation are tightly coupled, even in present-day
galaxies; and b) that the evolution of the AGN luminosity function
documented in recent optical and X-ray surveys is driven by a decrease
in the characteristic mass scale of actively accreting black holes.

\end{abstract}




\section{Introduction}

Over the past few years there have been remarkable developments in our
understanding of active galactic nuclei (AGN) and their role in galaxy
formation and evolution.  There is now compelling evidence (Miyoshi et
al. 1995; Genzel et al. 2000) for the existence of the supermassive
black holes that were long hypothesized as the power-source for active
galactic nuclei (Salpeter 1964; Lynden-Bell 1969). The local mass
density in these black holes is sufficient to have powered the total
emission of the AGN population over cosmic time
if the accreting material is assumed to
radiate with an efficiency near the upper end of the plausible range
(Yu \& Tremaine 2002; Marconi et al. 2004).  The tight correlation
between the mass of the black hole and the velocity dispersion and
mass of the galactic bulge within which it resides (Ferrarese \&
Merritt 2000; Gebhardt et al. 2000; Marconi \& Hunt 2003) is
compelling evidence for a close connection between the formation of
the black hole and that of its host galaxy (e.g. Cattaneo et al. 1999;
Kauffmann \& Haehnelt 2000; Granato et al. 2001). Finally, deep
x-ray surveys have recently
established that the AGN population exhibits so-called ``cosmic
down-sizing'': the space density of AGN with low x-ray luminosities
peaks at {\em lower redshift} than that of AGN with high x-ray
luminosities (Steffen et al 2003; Ueda et al. 2003).  These results
indicate that a substantial amount of the total black hole growth has
occured more recently than would have been deduced based on optical
surveys of powerful quasars (Boyle et al. 2000; Fan et
al. 2001). However, from these surveys alone it is not clear whether
this downsizing in luminosity is driven by a decrease of the accretion
rate in Eddington units, or by a decrease in the characteristic
mass-scale for the actively growing black holes. Breaking this
degeneracy requires information about black hole mass.

The co-evolution of galaxies and black holes can be directly
investigated through deep surveys spanning a broad range in redshift.
In the present paper, however, we take a complementary approach and
use the Sloan Digital Sky Survey (SDSS) to examine the relationship
between galaxies and their supermassive black holes in the present-day
Universe.  This paper builds on the results in Kauffmann et
al. (2003a: hereafter K03), in which we identified a sample of nearly
23,000 narrow-line AGN in the SDSS and investigated the properties of their host
galaxies as a function of AGN luminosity. We found that AGN reside
almost exclusively in massive galaxies with structural properties
similar to normal early-type systems. We also found that more powerful
AGN are hosted by galaxies with younger mean stellar ages.  We
interpreted the relation between AGN luminosity and stellar age as
saying that if gas is available to fuel the central supermassive black
hole, it also forms stars at the same time.  In the present paper we
seek to understand the implications of these results more deeply by
transforming from observables such as AGN luminosity and stellar
velocity dispersion, to ``physical'' quantities such as accretion rate
and black hole mass.

In section 2 we briefly review how we use the SDSS spectra and images
to measure the key properties of the black holes and of the galaxy
population that hosts them.  In section 3 we study the properties of
black holes that are accreting at the present day, in particular their
distributions of mass and of accretion rate in Eddington units. We
also study the properties of the host galaxies in which black holes
are currently growing, concentrating on their mass, density, structure
and star formation rate. In section 4 we conclude with a brief
discussion of our results.

\section{Methodology}

\subsection{The SDSS Data}

The data analyzed in this paper are drawn from the Sloan Digital Sky
Survey (York et al. 2000; Stoughton et al. 2002). The SDSS is using a
dedicated 2.5-meter wide-field telescope at the Apache Point
Observatory to conduct an imaging and spectroscopic survey of about a
quarter of the sky. The imaging is conducted in the $u$, $g$, $r$,
$i$, and $z$ bands (Fukugita et al. 1996; Smith et. al. 2002) using a
drift scan camera (Gunn et al. 1998). The images are calibrated
photometrically (Hogg et al. 2001) and astrometrically (Pier et
al. 2003) and are used to select galaxies (Strauss et al. 2002).
These are observed with a pair of multi-fiber spectrographs built by
Alan Uomoto and his team, with fiber assignment based on an efficient
tiling algorithm (Blanton et al. 2003).  The data are all
spectrophotometrically calibrated using observations of subdwarf F
stars in each 3-degree field (Tremonti et al., in prep).  Our sample
is the same as that described in K03. It is based on spectra of
122,808 galaxies with 14.5 $< r <$ 17.77 in the ``main'' galaxy sample
in the publically released SDSS Data Release One (DR1; Abazaijan et al
2003).

Our methodology is described in detail in K03, but we briefly
summarize it here for the convenience of the reader. We have written
special purpose code that fits each SDSS galaxy spectrum with a
stellar population model (Bruzual \& Charlot 2003) and then subtracts
this model to leave a pure emission-line spectrum (Tremonti et al., in
prep).  Following Baldwin, Phillips, \& Terlevich (1981; hereafter
BPT) and Veilleux \& Osterbrock (1987), we have identified AGN based
on the flux ratios of two pairs of the strongest narrow emission-lines
in their optical spectra: [OIII]$\lambda$5007/H$\beta$ {\it vs.}
[NII]$\lambda$6584/H$\alpha$ (henceforth the ``BPT diagram'').  Since
our aim is to relate the properties of the AGN to those of their host
galaxies, we have excluded objects (Type 1 Seyferts and quasars) in
which the AGN themselves contribute significantly to the continuum in
the SDSS spectra or images. Using the criteria adopted by K03 leads to
a sample of 22,623 narrow-line (Type 2) AGN.

In K03 we found the fraction of galaxies containing weak AGN to
decrease with redshift. This is because the projected size of the SDSS
3 arcsec diameter fibers increases with redshift (from $\sim$ 3 kpc to
12 kpc over the range $z \sim$ 0.05 to 0.2).  Dilution of AGN emission
by increasing amounts of galaxy light then causes us to miss more and
more weak systems with increasing redshift.  As discussed below, in
the present paper we will use the [OIII]$\lambda$5007 emission-line as
a proxy for AGN power. The AGN in our sample span a wide range in [OIII]
luminosity (from roughly $10^{5}$ to $10^{8.5} L_{\odot}$).
\footnote{We use $H_0$ = 70 km s$^{-1}$
Mpc$^{-1}$, $\Omega_M$ = 0.3, and $\Omega_{\Lambda}$ = 0.7 throughout this
paper.}
Based on the tests described in K03, we find that 
our sample is complete for AGN with [OIII] luminosities greater than
$10^6 L_{\odot}$ (we recognize these as AGN even at the outer limit of the
survey volume).
\footnote{Note that in K03 we used the Balmer decrement to correct the [OIII]
luminosity for dust extinction. We do {\it not} make these corrections
in the present paper because it complicates the empirical assessment of
bolometric corrections, as discussed below.}

\subsection{Derived Galaxy Properties}

As described in Kauffmann et al. (2003b) and Brinchmann et al.  (2004
- hereafter B04) we have used the SDSS spectra and images to derive a
set of fundamental physical properties for each galaxy. We summarize
these now.

We have used two age-sensitive spectral features to characterize the
stellar population: the 4000\AA\ break $D_n(4000)$ and the H$\delta$
absorption-line ($H\delta_A$).  These indices are compared with a
large grid of model galaxy spectra to determine the near-IR (SDSS
$z$-band) stellar mass-to-light ratio. A comparison of model and
observed spectral energy distributions yields a measure of the
attenuation of the $z$-band luminosity by dust.  These parameters are
then used to determine the stellar mass ($M_*$) from the $z$-band
Petrosian magnitude.
\footnote{All stellar masses and star formation rates are based on a Kroupa
(2001) initial mass function.}
We also use the Petrosian half-light radius in the $z$-band image 
($R_{50}(z)$) to measure the effective surface mass density 
$\mu_* = 0.5M_*/(\pi R_{50}(z))^2$.

Star formation rates are normally derived
for the region sampled by the spectroscopic fiber (SFR$_{{\rm
fib}}$) from the extinction-corrected luminosities of the most
prominent emission-lines. These lines cannot be used for galaxies with
AGN, and SFR$_{{\rm fib}}$ is instead estimated from D$_n$(4000),
based on the strong correlation between $D_n(4000)$ and specific
star-formation rate $SFR/M_*$ found for galaxies without AGN.  The
correction from SFR$_{{\rm fib}}$ to total star formation rate
(SFR$_{{\rm tot}}$) is made using the $g-r$ and $r-i$ color of the
galaxy in the region outside the fiber. This method exploits the fact
that for normal galaxies, the values of these colors measured inside
the fiber are a reasonably good predictor of $SFR_{{\rm
fib}}/L_{i,{\rm fib}}$ (see B04 for more details).  The
fiber typically encloses 20\% to 50\% of the galaxy light.

We use the SDSS concentration index $C = R_{90}/R_{50}$ (defined as
the ratio of the radius enclosing 90\% of the total $r$-band flux to
that enclosing half the $r$-band flux) as a measure of galaxy
structure. Shimasaku et al. (2001) and Strateva et al. (2001) show
that there is a good correspondence between $C$ and Hubble type, with
$C \sim$ 2.6 marking the boundary between early-type galaxies (E,S0
and Sa) and late-type systems (Sb-Irr) .

\subsection{Derived AGN Properties}

In this section we discuss how we use the SDSS data to estimate the
AGN bolometric luminosity.  First we motivate our use of the
[OIII]$\lambda$5007 emission line luminosity and describe our
procedure for correcting it for any contribution from star
formation. We then discuss the empirically-based bolometric correction
that is applied to the [OIII] luminosity.

\subsubsection{The [OIII] Luminosity}

We use the luminosity of the [OIII]$\lambda$5007 emission-line
($L_{O3}$) as a tracer of AGN activity.  This is usually the strongest
emission-line in the optical spectra of Type 2 AGN, and it is
significantly less contaminated than the others by contributions from
star-forming regions. This is important, because the relatively large
projected aperture of the SDSS spectra means that such contamination
is inevitable.

We have implemented the following procedure to statistically correct
$L_{O3}$ in each AGN for the contribution from star formation. First,
we divide the sample of all emission-line galaxies in K03 into three
classes, based on location in the BPT diagram (see Figure 1 in
K03). AGN-dominated spectra are defined to lie above the demarcation
line recommended by Kewley et al. (2001). Normal star forming galaxies
are defined to lie below the demarcation line adopted by K03.
``Composite'' objects where both an AGN and star formation contribute
significantly to the overall emission line spectrum lie between these
two lines. We extract all the AGN-dominated and normal star-forming
galaxies with stellar masses $M_* > 10^{10} M_{\odot}$.  Using these
two samples, we then populate the composite region of the BPT diagram
by adding the observed line luminosities of each of the AGN in turn to each
star-forming galaxies.  This creates a population of synthetic objects
with $L_{O3}$ in the range that covers the
region of the BPT diagram containing the real composite galaxies.  We
then bin the synthetic objects according to $L_{O3}$ and their
locations in the BPT diagram. We compute the {\em average} fractional
AGN contribution to $L_{O3}$ for all the objects in each bin and we
use this to correct $L_{O3}$ for the real composite galaxies that fall
in the same bin.  The AGN contribution to $L_{O3}$ ranges from 50\% to
90\% for the composite objects and is greater than 90\% for the
AGN-dominated objects.

\subsubsection{The Bolometric Correction}

In the standard paradigm (e.g. Antonucci 1993), AGN can be broadly
classified into two groups. In Type 1 objects the central black holes
and their associated continuum and broad emission-line regions are
viewed directly. In Type 2 objects these regions are obscured along
the observer's line-of-sight by a dusty torus which intercepts the
optical and ultraviolet emission and reradiates it in
the infrared.  The [OIII]$\lambda$5007 emission line arises in gas
which is photo-ionized by AGN radiation escaping along the polar axis
of the torus, but lying well outside the torus. The observed flux is
therefore little affected by viewing angle. 

We have included only Type 2 AGN in our analysis, because for these
objects the AGN itself has no significant impact on our estimation
either of the galaxy parameters described above (see K03), or of the
black hole mass (see below). Unfortunately the bolometric correction
to $L_{O3}$ for AGN can only be directly determined for Type 1 AGN,
where the clear view of the AGN allows the full spectral energy
distribution to be measured directly.  As just noted, in the standard
unified AGN model the [OIII]$\lambda$5007 emission-line provides a
good indication of the intrinsic luminosity of the AGN in both Type 1
and Type 2 systems. This can be tested by comparing the ratio of the
[OIII] flux to the AGN continuum flux for Type 1 and Type 2 AGN in
those wavelength regimes where absorption by the torus is least
significant and the contribution by starlight is minimal (mid/far
infrared, radio, and hard X-ray regimes).

Keel et al. (1994) found that the distributions of the ratio of the
infrared (25 and 60 $\mu$m) to [OIII] fluxes were identical in
complete samples of Type 1 and Type 2 AGN selected from the IRAS
survey. Similar results were found by Mulchaey et al. (1994) and
Heckman (1995) for independent samples of AGN. The ratio of nonthermal
radio continuum to [OIII] is the same in Type 1 and Type 2 AGN to
within a factor of two in both radio-quiet (Heckman, 1995) and
radio-loud AGN (Simpson 1998). Finally, Mulchaey et al. (1994) showed
that the mean ratio of hard X-ray to [OIII] flux is the same in Type 1
and Type 2 AGN. This was later verified by Bassani et al. (1999), so
long as Type 2 AGN with Compton-thick tori were excluded.  Thus, we
will assume that the bolometric correction to $L_{O3}$ is the same on
average in Type 2 and in Type 1 AGN.

We have determined the bolometric correction to $L_{O3}$ for Type 1
AGN in a two step process. First, we determine the mean ratio of 
the optical continuum and [OIII] luminosities using the extensive
SDSS samples of Type 1 Seyfert nuclei and low-redshift quasars
described in Zakamska et al. (2003) and K03. We have merged these two
samples and then restricted them to objects with $z <$ 0.3 (our Type 2
AGN have $<z> \sim$ 0.1). There is a strong linear correlation between
$L_{O3}$ and optical continuum luminosity (see Zakamska et al.  2003),
and we find an average ratio $L_{5000}/L_{O3} \sim$ 320 (where
$L_{5000}$ is the monochromatic continuum luminosity
$\lambda$$P_{\lambda}$ at 5000 \AA\ rest-frame). The dispersion
about this mean is 0.34 dex (one $\sigma$). 
We have confirmed these
results using a much smaller sample of nearby Type 1 Seyfert nuclei
($<z> \sim$ 0.03) compiled by Dahari \& De Robertis (1988). 

The second step is to then determine the mean ratio of the bolometric
and optical continuum luminosities in Type 1 AGN.
The mean Type 1 AGN intrinsic spectral energy
distribution in Marconi et al. (2004) gives $L_{Bol}/L_{5000}$ =
10.9. Elvis et al (1994) find a variance in this ratio of 0.16 dex
among Type 1 AGN. The implied
bolometric correction is then $L_{Bol}/L_{O3} \sim$ 3500, with
a variance of 0.38 dex. We have verified these results using a smaller sample
of 28 Type 1 AGN in the Dahari \& De Robertis (1988) catalog
with good mid-IR, optical, UV, x-ray, and [OIII] fluxes. While the
bolometric correction to $L_{O3}$ will be uncertain by $\pm$0.38 dex
for any individual AGN, we will be performing integrals over
a sample of more than 20,000 AGN, so we need only have a good  estimate
of the {\it average} bolometric correction. 

It is important to note that the samples used above to calibrate the
bolometric correction are dominated by relatively powerful AGN with
$L_{O3} \sim 10^{6.5}$ to $10^{9} L_{\odot}$.  Since our SDSS sample
extends to lower luminosities, we have examined these
samples to look for a systematic luminosity dependence in the
bolometric correction. Although we find no evidence for such a trend,
in the remainder of the paper we will present results based both on
our full sample of AGN (all $L_{O3}$) and on the strong AGN alone
($L_{O3} \geq 10^{6.5} L_{\odot}$). This luminosity cut also
effectively excludes the LINER class of Type 2 AGN. Such AGN may well
have very different overall spectral energy distributions than the
more powerful Seyfert nuclei and quasars (e.g. Ho 2004). 
Finally,  while $L_{O3}$ is a reasonable proxy for
$L_{bol}$ in the unified model, it does depend directly on
the availablity of gas clouds to ionize. Thus, there could be 
correlations between host galaxy properties and $L_{bol}/L_{O3}$.

\subsubsection{Missing Type 1 AGN}

As explained above, our sample explicitly excludes Type 1 AGN. In this
paper, we are interested in documenting which black holes are growing
and where they are growing. Thus, it is important to keep in mind the
limitations resulting from this exclusion.

In the simplest form of the unified model, the Type 1 and Type 2 AGN
are drawn from the same parent population of galaxies and differ only
in the orientation of the observer's line-of-sight. If all tori were
similar, our results for black hole demographics could simply be
scaled upward by the inverse of the Type 2 AGN fraction. The real
situation is more complex. Using the same definition of Type 2 AGN as
K03, Hao (2003) and Hao et al. (in preparation) find that the fraction of
Type 2 AGN in the
SDSS is a weak function of AGN luminosity, decreasing from $\sim60\%$
to $\sim30\%$ over the range $L_{O3} \sim 10^{5.5}$ to $10^{8.5} L_{\odot}$.
The unified model then implies that the torus opening angle is a
weakly increasing
function of AGN luminosity. The implied total volume-averaged
correction for missing Type 1 AGN would be a factor of $\sim 2$. 

\subsection{Derived Black Hole Properties}

We use the observed correlation between black hole mass $M_{BH}$ and
bulge velocity dispersion $\sigma_*$ (Tremaine et al. 2002)
\begin {equation}
\log M_{BH} = 8.13 + 4.02 \log(\sigma_*/200 {\rm km s}^{-1}) 
\end{equation} 

to derive a black hole mass for galaxies both with and without AGN.
A velocity dispersion is
automatically measured for each SDSS galaxy by fitting the observed
spectrum with a linear combination of galaxy template spectra
broadened by a Gaussian kernel (Schlegel et al. in preparation). The
instrumental resolution of the SDSS spectra is $\sigma_{instr} \sim$
60 to 70 km s$^{-1}$, so we restrict our analysis to galaxies with
measured values $\sigma_* \geq$ 70 km s$^{-1}$ (corresponding to log
$M_{BH} \geq$ 6.3). 
We do not apply any aperture corrections to the stellar velocity
dispersions, because such corrections are small in early-type galaxies
in the SDSS (Bernardi et al. 2003).

For the galaxies without an AGN, we
restrict our estimates of black hole mass
to galaxies with stellar surface mass densities $\mu_* > 3 \times 10^8
M_{\odot} $ kpc$^{-2}$.  As shown by Kauffmann et al. (2003c), this
value of $\mu_*$ marks the point where the galaxy population undergoes
an abrupt transition from disk-dominated galaxies with young stellar
populations and ongoing star formation to bulge-dominated
``early-type'' galaxies. In disk-dominated galaxies, the velocity
dispersion measured inside the fiber will not provide an accurate
estimate of black hole mass.  Note that if we do include galaxies with
$\mu_* < 3 \times 10^8 M_{\odot}$ kpc$^{-2}$, the integrated mass
density in black holes with $6.3 < \log M_{BH} < 7$ would increase by
a factor of $\sim 2$. The integrated mass density in more massive
black holes would only increase by $\sim 1.2-1.4$. None of the
qualitative conclusions presented in this paper would change.
For AGN hosts, we derive a black hole mass even if                             
$\mu_* <  3 \times 10^8 M_{\odot} $ kpc$^{-2}$ (since
the AGN  indicates that there is a  black hole). As we will show in
section 3.2 below, AGN in host galaxies with $\mu_* < 3 \times 10^8 M_{\odot}$ kpc$^{-2}$
contribute very little 
to the volume-averaged black hole mass and accretion rates.

Using the bolometric luminosity,
we calculate a mass accretion
rate $\dot{M}$ assuming a fiducial value of $\epsilon$ = 10\% for the radiative
efficiency (e.g. Yu \& Tremaine 2002). 
We then use the black hole mass to calculate an Eddington luminosity
and associated Eddington ratios
($L_{bol}/L_{edding}$ = $\dot{M}_{BH}/\dot{M}_{edding}$ for our assumed
fixed value of $\epsilon$).
On physical grounds, it is likely that the values we have adopted for both the 
bolometric correction to $L_{03}$
and $\epsilon$ will not be valid at sufficiently
low values
of accretion in Eddington units (Narayan et al. 1998; Ho 2004).
Indeed we see a transition 
in AGN type from Seyfert to LINER at a value $L_{bol}/L_{edding} \sim$ 1\%
in our sample. Thus, 
in the remainder of the paper we will present results based both on
our full sample of AGN and on only the AGN with 
$L_{bol}/L_{edding} \geq 10^{-2}$. Clearly, if substantial growth in
the mass of the present-day black hole population is occuring
via accretion with a very low radiative efficiency, we will miss this.
However, as  discussed by Yu \& Tremaine (2002) and  Marconi et al. (2004),  
the local mass
density in present-day black holes is sufficient to have powered the total
emission of the AGN population over a Hubble time only if the accreting 
material is assumed to
radiate with $\epsilon$ no smaller than the value we have adopted.
The overall
contribution to black hole growth by accretion with low radiative efficiency
has evidently not been very important.

\section{Results}

In this section we study where accretion onto black holes, as traced
by [OIII] emission from Type 2 AGN, is occurring at the present day. As
discussed by Ho et al (1997) and K03, 40--50\% of massive galaxies
have detectable AGN, but [OIII] emission is very weak in most of these
systems so their black holes are clearly not growing very rapidly.
Because powerful AGN are rare (only 5\% of $10^{11} M_{\odot}$
galaxies have Type 2 AGN with log $L_{O3}>10^{6.5} L_{\odot}$), it is more
meaningful to work with {\em volume-averaged} quantities than to
present results for individual AGN. Both normal galaxies and AGN in
our sample are selected from an $r$-band magnitude-limited redshift
survey. We can thus follow the standard procedure and weight each
galaxy or AGN by $1/V_{max}$, where $V_{max}$ is the volume over which
the object would have been detectable (Schmidt 1968;  see K03 and
Kauffmann et al 2003b
for more details). In this way, we
are able to compute the integrated [OIII] luminosity from Type 2 AGN per unit
stellar mass or per unit black hole mass. Except when noted otherwise,
we will perform these volume integrals over {\it all} galaxies, not
just the AGN hosts. 

Throughout the rest of the paper we will denote these volume-integrated
quantities by scripted symbols to distinguish them from values for
individual objects (e.g. $\cal M_{\rm BH}$ vs. $M_{BH}$ and
$\cal L_{\rm O3}$ vs. $L_{O3}$).

\subsection{Which Black Holes are Growing?}

In the left hand panel of Fig.~1, the long-dashed curve shows how
$\cal M_{\rm BH}$, the integrated mass contained in black holes in our
sample, is distributed. We have checked that the distribution we compute
 agrees well with several
other recently derived black hole mass functions
(see Marconi et al (2004), and references therein).
Recall that we can estimate black hole mass
for galaxies with stellar velocity dispersions larger than 70 km/s,
which corresponds to $M_{BH} =10^{6.3} M_{\odot}$. As can be seen, the
bulk of the mass resides in black holes with masses between $\sim10^{7.5}$
and $10^{8.5} M_{\odot}$. However, the bulk of the Type 2 AGN 
luminosity is produced by black holes with significantly lower masses.
This is illustrated by 
the solid, dotted, dashed and dashed-dotted curves, which show how the
integrated [OIII] luminosity from Type 2 AGN ($\cal L_{\rm O3}$) is
partitioned among black holes of different mass.  The dashed line
shows the result that is obtained if we integrate over these AGN
without correcting the [OIII] luminosities of composite AGN for the
contribution from star formation.  The solid line shows the result if
this correction is included using the procedure described in section
2.3.1.  The dotted line shows what happens if we only consider AGN
with log $L_{O3}> 10^{6.5} L_{\odot}$ {\em and} we correct the [OIII]
luminosities for star formation. The dashed-dotted line shows the result
for AGN with $\dot{M}_{BH}/M_{edding} > 10^{-2}$ and correction
for star formation.  As can be seen, 
uncertainties in the nature of the bolometric correction have 
only a small effect on our results.  The peak contribution to
$\cal L_{\rm O3}$ comes from black holes with masses $\sim 3 \times
10^7 M_{\odot}$, which is a factor of 3 less than mass at which
the contribution to the black hole mass inventory peaks.

The right hand panel of Fig. 1 shows how the integrated [OIII]
luminosity from our sample of Type 2 AGN is distributed as a function
of $L_{O3}$. 
Above $ 10^6 L_{\odot} $, $\cal L_{\rm O3}$ is spread
rather evenly between low-luminosity and high-luminosity systems up to
$L_{O3} \sim 10^8 L_{\odot}$. The fall-off below
$ 10^6 L_{\odot} $ is caused by incompleteness in our
sample of weak AGN at higher redshifts in the SDSS (see Kauffmann et al (2003c)
for a detailed discussion). We have examined
a low-redshift subset our AGN sample ($z <$ 0.04) in order to estimate the effect
of this imcompleteness. Our tests show
show that low luminosity objects ($L_{O3} = 10^5$ to $10^6 L_{\odot} $) 
that are missing from the
the full sample plotted in Fig.1 would account for 16\% of the total
volume-integrated [OIII] emission by all Type 2.  We are thus
slightly underestimating $\cal L_{\rm O3}$.

In Fig.~2, we plot $\cal L_{\rm O3}$/$\cal M_{BH}$ as a function of
$\log M_{BH}$. If we convert $\cal L_{\rm O3}$ into a volume-averaged
accretion rate using the relations described in section 2, we can
recast $\cal L_{\rm O3}$/$\cal M_{BH}$ as a growth time for the
population of black holes ($\cal M_{BH}$/$\dot{\cal M}_{BH}$).  Fig. 2
shows that the most rapidly growing black holes are those with $M_{BH}
<$ few $\times 10^7 M_{\odot}$, with an implied growth time of only 
$\sim$20 to 40 Gyr (twice the age of the universe). Above this mass, the
growth time increases very rapidly to become orders
of magnitude longer than the Hubble time.
Alternatively, we can think of
$\dot{\cal M}_{BH}$ $t_{Hubble}$/$\cal M_{BH}$ as a
measure of the ratio of the current to the past-averaged growth rate
of black holes. Our results indicate that for low-mass black holes,
this is around a half, just for growth associated with Type 2 AGN alone.
In contrast, the population of more massive black holes ($> 10^8
M_{\odot}$)
must have formed at significantly higher redshifts, as 
they are currently experiencing very little additional growth.

One might ask whether $\cal L_{\rm O3}$/$\cal M_{BH}$ is
systematically overestimated for low mass black holes because we have
excluded low-density, disk-dominated galaxies in our estimates of
integrated black hole mass. We have tested the importance of this
effect by placing a black hole with mass given by equation (1) in all
galaxies with $\sigma > 70$ km s$^{-1}$, irrespective of stellar
surface density. We find that this only increases the growth times of
the lowest mass black hole population in our sample by a factor of 1.6.

As well as looking at the integrated accretion onto black holes of
given mass, it is also interesting to study the {\em distribution} of
accretion rates in these systems. This is shown in Fig.~3. We
normalize the accretion rate by dividing by the Eddington rate for
each object.  The left panel shows the cumulative fraction of black
holes that are accreting above a given rate. Results are shown for
different ranges in black hole mass, with mass increasing from the
solid curve on the right ($3 \times 10^6 M_{\odot}$) to the dotted
curve on the left ($10^9 M_{\odot}$) by a factor of 3 in each case.
Note that the sample has been restricted to AGN with 
$\log L_{O3}>10^{6.5} L_{\odot}$ (with a correction for the contribution
to $L_{O3}$ from star formation). Thus, we are able to go further down in
$\dot{M}_{BH}/ \dot{M}_{edding}$ for more massive black holes.  Fig.~3
shows that the accretion rate functions cut off fairly neatly at
$\dot{M}_{BH}/\dot{M}_{edding} \sim 1$. This gives us confidence that
our conversions from $L_{O3}$ to accretion rate and from $\sigma$ to
black hole mass are yielding reasonable answers.  There are more low
mass black holes than high mass black holes with very large accretion
rates near the Eddington limit. Roughly 0.5 percent of black holes
with $M_{BH} = 10^7 M_{\odot}$ are accreting above a tenth the
Eddington limit. For black holes with $M_{BH} \sim 3 \times 10^8
M_{\odot}$, the fraction accreting above a tenth Eddington has dropped
to $10^{-4}$. This explains the exceptionally rapid decline (Figure 2) 
in $\cal L_{\rm O3}$/$\cal M_{BH}$ above $\log M_{BH} \sim$ 8.6
for the sample in which we exclude AGN with $L_{bol}/L_{edding} <$ 1\% 
(the dot-dash line style). 

The right panel of Figure 3 shows the cumulative fraction of the total accreting
mass (as traced by Type 2 AGN with log $L_{O3} > 10^{6.5} L_{\odot}$) as a
function of $\dot{M}_{BH}/\dot{M}_{edding}$. Results are shown for low
mass black holes with $M_{BH} < 3 \times 10^7 M_{\odot}$ (solid) and
for high mass black holes with $M_{BH} > 3 \times 10^7 M_{\odot}$
(dashed).  As can be seen, much of the black hole growth is occurring
in systems with high Eddington ratios.  For low mass black holes, 50\% of
all growth occurs in AGN that are within a factor of $\sim$5 of the
Eddington luminosity. For high mass black holes, half of the
present-day growth occurs in AGN radiating above $\sim$8\%
Eddington.  Yu \& Tremaine (2002) have shown that the mass density in
black holes estimated from integrating the luminosity function of
quasars over all cosmic epochs agrees rather well with the total mass
density in black holes in galaxies at the present day.  Their analysis
did not include low-luminosity or Type 2 AGN and it has thus been
something of a puzzle why these two estimates of black hole mass
density should agree so well.  Our analysis demonstrates that although
low-luminosity AGN are numerous, their contribution to the growth of
black holes is not dominant.

\subsection{Where are Black Holes Growing?}

In Fig.~4, we plot the integrated [OIII] luminosity from Type 2 AGN binned up
as a function of stellar mass $M_*$, of stellar surface mass density
$\mu_*$, of concentration index $C$ and of 4000 \AA\ break strength
D$_n$(4000).  Fig.~4 shows that most present-day accretion is taking
place in galaxies with young stellar ages (D$_n$(4000)$< 1.6$),
intermediate stellar masses ($10^{10}$ -few $\times 10^{11}
M_{\odot}$), high surface mass densities ($3 \times 10^8 - 3 \times
10^9$ $M_{\odot}$ kpc$^{-2}$), and intermediate concentrations ($C
\sim$ 2.2 to 3.0).  K03 already showed that powerful Type 2 AGN occur in
galaxies with young stellar populations, so it is no surprise that
most of the [OIII] emission comes from galaxies with low values of
D$_n$(4000).  It is quite remarkable, however, that the [OIII]
emission from AGN peaks so near the {\em transition values} of $M_*$,
$C$ and $\mu_*$ where the galaxy population switches abruptly from
young, star-forming and disk-dominated to old, quiescent and
bulge-dominated (Kauffmann et al 2003c).

\subsection {Bulge building: the relation between star formation 
and accretion onto black holes}

In section 3.2, we showed that the population of
low mass black holes is still growing
rapidly (at one half of the past-averaged rate, 
even ignoring the contribution
of Type 1 AGN). If the very
tight relation between black hole mass and bulge mass is to be
maintained, the host bulges of these systems must be ``forming'' at a
comparable rate.

In Fig. 5, we plot $\cal SFR$/$\dot{\cal M}_{\cal BH}$, the ratio of
the integrated star formation rate in {\it all} galaxies to the
integrated accretion rate onto black holes, as traced by [OIII]
emission from Type 2 AGN.
The thick black line shows the result if one considers only
the SFR inside the SDSS fiber aperture.  The thin black line shows the
result using our estimates of {\em total} SFR (see section 2.2).  The
dashed lines include all AGN, but with the contribution to
$L_{O3}$ from star formation removed. The 
solid lines show what happens when the sample is restricted to AGN
with $L_{O3}> 3 \times 10^6 L_{\odot}$. We have chosen to plot $\cal
SFR$/$\dot{\cal M}_{\cal BH}$ as a function of black hole mass
$M_{BH}$ and of stellar surface mass density $\mu_*$. 

From Fig. 5 it is
clear that black hole growth is closely linked to star formation in
the bulge.  At low values of $\mu_*$ characteristic of disk-dominated
galaxies, the ratio of $\cal SFR$/$\dot{\cal M}_{\cal BH}$ rises
steeply. This is because very few of these galaxies host AGN, but
there is plenty of star formation taking place in galaxy disks.  At
values of $\mu_*$ above $3 \times 10^8 M_*$ kpc $^{-2}$ , $\cal
SFR$/$\dot{\cal M}_{\cal BH}$ remains roughly constant.  Moreover, its
value is $\sim 1000$, which is in good agreement with the
empirically-derived ratio of bulge mass to black hole mass (Marconi \&
Hunt 2003).  Note that the star formation rate estimated within the
fiber is likely to underestimate the true star formation rate in the
bulge, particularly for the largest and most massive galaxies.
Conversely, the total SFR is probably an overestimate, because many
galaxies will have star-forming disks. Given the uncertainties in the
bolometric correction, in the conversion from $L_{{\rm bol}}$ to black
hole accretion rate, in the effect of the missing Type 1 AGN, and in our
estimates of SFR, we find
it remarkable that $\cal SFR$/$\dot{\cal M}_{\cal BH}$ comes out
within a factor of a few of the value that is expected from the
$M_{BH}$--$M_{bulge}$ relation.

Finally, in Fig.~6 we compare the average growth time of 
the population of supermassive
black holes with the average growth time of 
the population of galactic bulges. For
galaxies with stellar masses in the interval $(M_{*}, M_{*}+\Delta
M_{*})$ the black hole growth time is estimated by calculating the
total mass in black holes ($\cal M_{BH}$) and dividing by the
accretion rate deduced from the total Type 2 AGN luminosity in [OIII] ($\cal
L_{\rm O3}$). We restrict this sample to AGN with
$L_{O3}> 3 \times 10^6 L_{\odot}$.
The growth time of the bulge is estimated from the
ratio of the total stellar mass within the fiber aperture in these
systems to the total star formation rate measured inside the fiber.
We perform these integrals only over the bulge-dominated normal galaxy
population ($\mu_* > 3 \times 10^8 M_{\odot}$ kpc$^{-2}$), but
include the contribution to black hole growth by all AGN. 
As can be seen, the black hole and bulge growth times track each other
remarkably closely for galaxies with $M_* > 10^{10} M_{\odot}$. 
For low mass bulges/black holes, the growth times are of order the Hubble
time, but they increase by more than an order of magnitude for massive
bulges/black holes. This is consistent with the fact that the smaller bulges
in later type galaxies tend to be younger than the larger bulges
in earlier type galaxies (Carollo 2004 and references therein).

\section{Discussion}

We have used a sample of $\sim$23,000 AGN, and a complete sample
of $\sim$123,000 SDSS galaxies from which they were drawn, to investigate the
accretion-driven growth
of supermassive black holes in the low-redshift universe, and to
relate this growth to the properties of the galaxy population.  Before
discussing our results, we remind the reader of the limitations of our
approach, and the assumptions we have made to deal with them.

\begin{enumerate}
\item
We have used optical emission-line ratios to identify our AGN. If
there is a substantial population of accreting black holes in SDSS
galaxies that we do not classify as AGN
(e.g. Maiolino et al. 2003), we will have missed their
contribution to the total black hole growth. 
\item
We have considered only Type 2 (narrow line) AGN in which the intense
radiation from the central accretion disk is completely obscured along
our line-of-sight. The effect of the missing Type 1 AGN is relatively
small (a factor of $\sim$2) for our global assessments, but the slow
increase in the ratio of Type 1 to Type 2 with increasing AGN
luminosity in the SDSS (Hao 2003) means that we systematically underestimate
the contribution of more powerful AGN by a factor of a few.
\item 
We have used the luminosity of the [OIII]$\lambda$5007 emission line
to estimate Type 2 AGN luminosity, assuming that the bolometric correction is
the same as for Type 1 AGN.
This approach is motivated by
the unified model for AGN, and has been empirically validated in the
literature. 
\item
We are complete only for Type 2 AGN with $L_{O3}$
above 10$^6 L_{\odot}$. We estimate this causes us to miss about 16\%
of the total volume-integrated [OIII]$\lambda$5007 luminosity from Type 2 AGN.
\item 
We have calculated black hole accretion rates for an assumed radiative
efficiency of 10\%. We will undercount accretion that occurs with
lower radiative efficiency.
\item
We have used the stellar velocity dispersion measured within the
central ($\sim 3$ to 10 kpc diameter) region to estimate the black
hole mass, using the $M_{BH}-\sigma$ relation. We only derive black
hole masses for bulge-dominated normal galaxies with $\mu_* > 3 \times 10^8
M_{\odot}$ kpc$^{-2}$, but include all the AGN hosts.
\end {enumerate}

Our most surprising result is that the growth time of the population of low mass
black holes ($<3 \times 10^7 M_{\odot}$) is short (only $\sim$ two Hubble
times, even ignoring the contribution from Type 1 AGN).
Although low mass black holes are growing substantially {\em
in the mean} at the present day, only 0.2\% of them are growing at the
rate which counts for 50\% of the fuelling. This implies that strong
fuelling events can only last $\sim 3 \times 10^7/N$ years if there
are $N$ events per black hole per Hubble time. 
This implies that the strong accretion phase is a
fraction of a characteristic galaxy dynamical time of $> 10^8$ yr, the shortest
timescale over which significant variations in the star formation rate
could occur.
The correlation
between the stellar age of the host galaxy and AGN luminosity found by K03
might then be a
consequence of a starburst occurring simultaneously with the
maximum accretion phase, followed by a fairly slow tail-off in
accretion as the stellar population ages.

By contrast, we find that massive black holes have very long growth
times. On average, black holes with masses of $\sim10^9 M_{\odot}$ are
currently accreting at a rate that is several orders-of-magnitude
lower than their past averaged rate. This population evidently formed
early and has evolved little at recent times. This is consistent with
inferred black hole masses for high-z quasars (e.g. McLure \&
Dunlop 2004).

We have also shown that the ratio of the volume-averaged star
formation rate to the volume-averaged black hole accretion rate is
$\sim10^3$ for bulge-dominated galaxies ($\mu_* > 10^{8.5} M_{\odot}$
kpc$^{-2}$). Given the uncertainties in our estimates
of the accretion rate, this is remarkably similar to the present ratio of
stellar to black hole mass in galaxy bulges.  For the population of
such galaxies with stellar masses greater than $10^{10} M_{\odot}$, the
mean growth time of the black hole and the mean growth time of the
surrounding bulge (measured inside the SDSS fiber aperture) are very
similar.  After volume averaging, the growth of black holes through
accretion and the growth of bulges through star formation are thus
related at the present time in the same way that they have been
related, on average, throughout cosmic history.  It seems likely that
the processes that established the tight correlation between bulge
mass and black hole mass are still operating in low redshift AGN.

These results suggest a picture in which star formation and black hole
growth have been moving steadily and in parallel to lower and lower
mass scales since a redshift of $\sim$2. This is consistent with the
long established fact that at low redshift more massive galaxies tend
to have older stellar populations. Surveys of high redshift galaxies
have found direct evidence for such an evolutionary pattern, dubbed
``cosmic downsizing'' by Cowie et al. (1996), in the star formation
properties of galaxies as a function of mass. More recently x-ray
surveys (Ueda et al. 2003; Steffen et al 2003) have shown that the peak in
the co-moving emissivity of the AGN population occurs at lower and
lower redshift for AGN of lower and lower luminosity.  These surveys
show that a surprisingly large fraction of the total growth of the
black hole population has occurred relatively recently in AGN of
modest luminosity. For example, Barger et al. (2003) find that over
half the x-ray background has been produced by accreting black holes
at $z <$ 1. 
However, these surveys do not establish whether this is
simply due to a decline with time in the average accretion rate (in
Eddington units), or rather to a decline in the characteristic mass
scale at which black hole growth occurs (true ``cosmic downsizing'').
Discriminating between these two possibilities requires estimates of
the black hole masses.  With the superb statistics provided by the
SDSS we have been able to show that real downsizing has occured in the
AGN population.

It is particularly intriguing that most black hole growth is currently
occurring in galaxies lying so near the values of galaxy mass,
density, and concentration where the galaxy population abruptly
transitions from low density, disk-dominated systems with ample
ongoing star formation to dense, bulge-dominated systems with little
star-formation. This may imply that today's AGN occur in a narrow
``habitable zone'' with a precipice to the low mass side (no black
holes) and an on-coming forest fire at higher masses (no cold
interstellar gas for fuel).  As we realize that the Era of AGN did not
end at high redshift, it seems that we must also admit that this
exciting epoch may be finally drawing to a close.



\acknowledgments

Funding for the creation and distribution of the SDSS Archive has been
provided by the Alfred P. Sloan Foundation, the Participating
Institutions, the National Aeronautics and Space Administration, the
National Science Foundation, the U.S. Department of Energy, the
Japanese Monbukagakusho, and the Max Planck Society. The SDSS web site
is http://www.sdss.org/.  The SDSS is managed by the Astrophysical
Research Consortium (ARC) for the Participating Institutions. The
Participating Institutions are The University of Chicago, Fermilab,
the Institute for Advanced Study, the Japan Participation Group, The
Johns Hopkins University, Los Alamos National Laboratory, the
Max-Planck-Institute for Astronomy (MPIA), the Max-Planck-Institute
for Astrophysics (MPA), New Mexico State University, University of
Pittsburgh, Princeton University, the United States Naval Observatory,
and the University of Washington.  SC thanks the Alexander von
Humboldt Foundation, the Federal Ministry of Education and Research,
and the Programme for Investment in the Future (ZIP) of the German
Government for financial support.
JB acknowledges the receipt of an ESA post-doctoral fellowship.




\pagebreak

\clearpage

\begin{figure}
\plotone{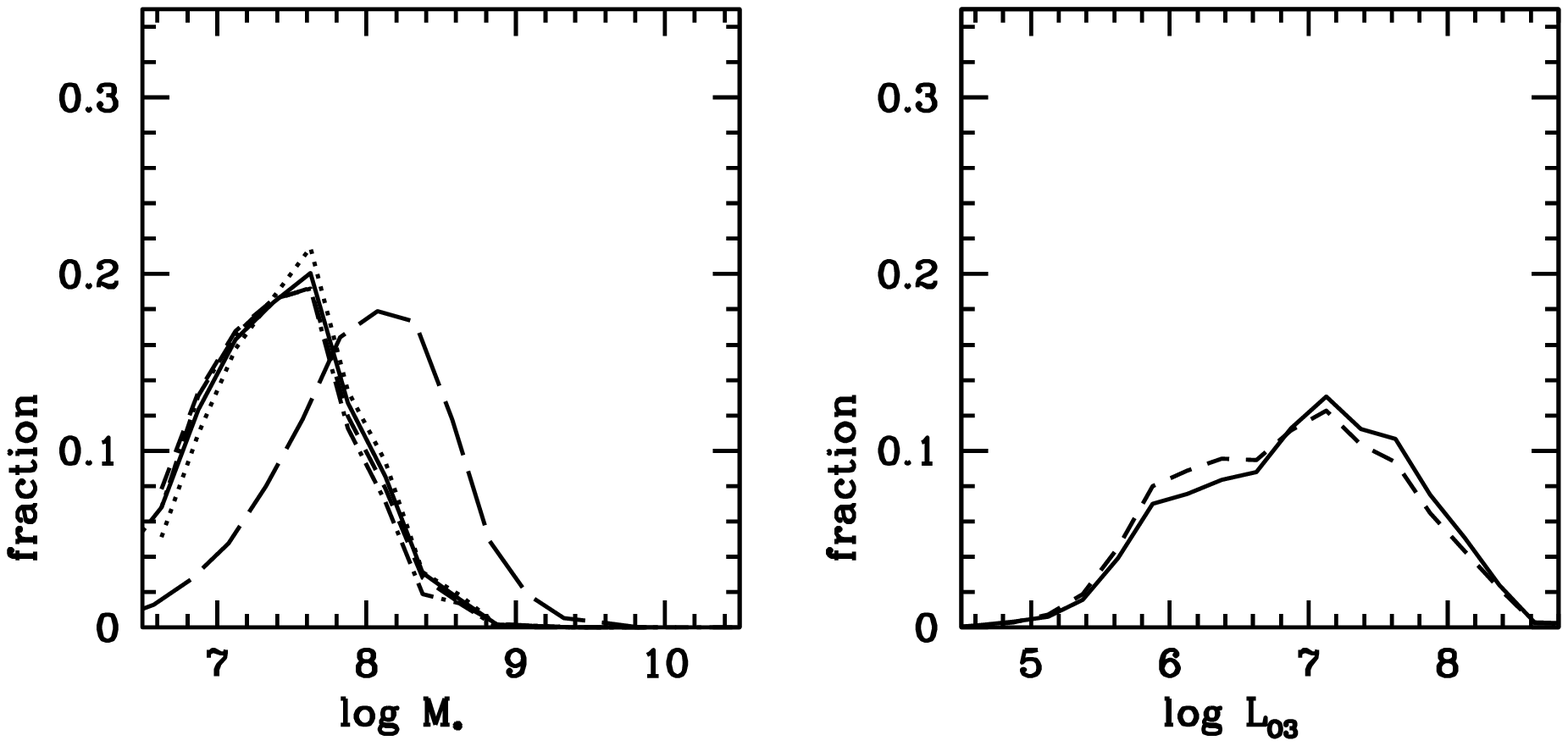}
\caption{Left: The long dashed curve shows the distribution over black
hole mass of the volume-weighted total mass of all black holes (active
and inactive) in our sample. The solid, dashed, dotted, and dash-dotted lines
show the distribution over black hole mass of the volume-weighted total
[OIII] luminosity from Type 2 AGN.  The dashed line shows the result obtained
by integrating over all AGN in the sample. The solid line shows the
effect of correcting the [OIII] luminosities for the contribution from
star formation. The dotted line shows the result obtained if one only
integrates over AGN with $\log$ $L_{O3} > 10^{6.5} L_{\odot}$ and one
also corrects $L_{O3}$ for star formation. The dashed-dotted line is for AGN
with $\dot{M}/M_{edding}> 10^{-2}$ and corrected for
star formation.  Right: The distribution
over $L_{O3}$ for individual AGN of the total volume-weighted [OIII]
luminosity from AGN. The solid line shows the result obtained when
$L_{O3}$ is corrected for star formation and the dashed line shows the
result without any correction. Note that the decline below
$\log$ $L_{O3} = 10^{6} L_{\odot}$ is due to incompletness in our sample.
\label{fig1}}
\end{figure}

\clearpage

\begin{figure}
\plotone{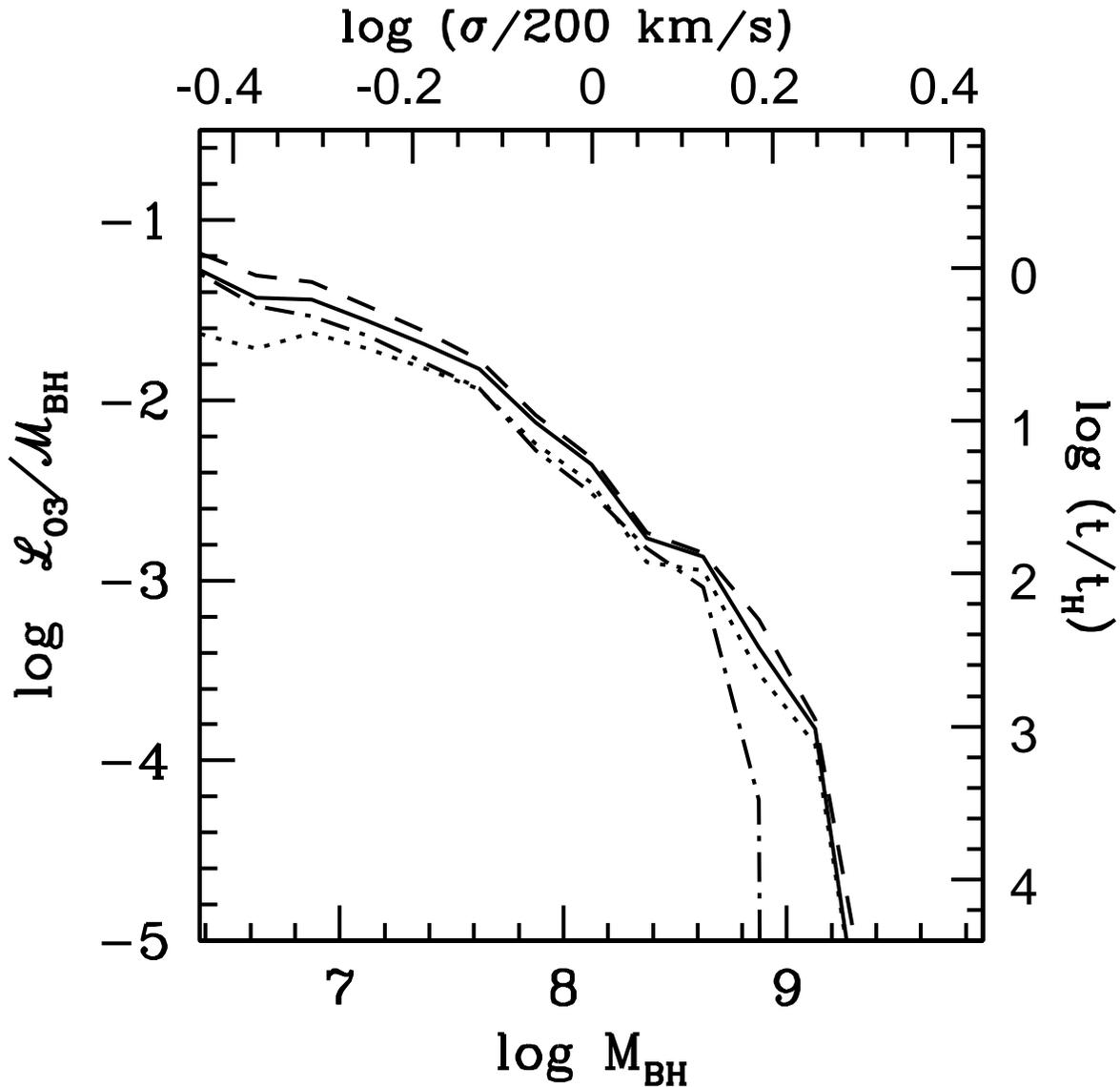}
\caption{The logarithm of the ratio of the total volume-weighted
[OIII] luminosity in Type 2 AGN to the total volume-weighted mass in black
holes (both in solar units) is plotted as a function of velocity
dispersion $\sigma$ (upper axis) and $\log M_{BH}$ (lower axis). The
right axis shows the corresponding growth time for the population of
black holes in units of the Hubble time.  The line types are as
described in Fig.~1.
\label{fig2}}
\end{figure}

\clearpage

\begin{figure}
\plotone{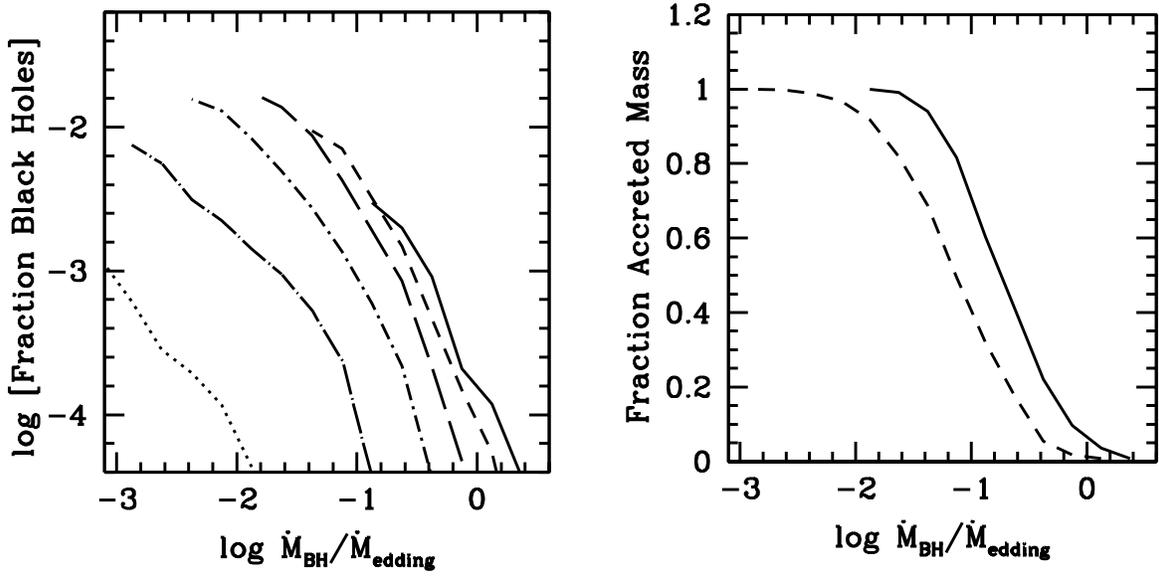}
\caption{Left: The fraction of all black holes accreting above a given
rate (in Eddington units).  Solid, short-dashed, long-dashed, short
dashed-dotted, long dashed-dotted and dotted curves are for black
holes with $3 \times 10^6$, $10^7$, $3 \times 10^7$, $10^8$, $3 \times
10^8$ and $10^9 M_{\odot}$, respectively.  Right: The cumulative
fraction of the total volume-weighted mass accretion rate contributed
by Type 2 AGN accreting above a given rate. The solid curve is for black
holes with $M_{BH} < 3 \times 10^7 M_{\odot}$ and the dashed curve is
for black holes with masses above this value. These distributions
are all computed for AGN with $\log$ $L_{O3} > 10^{6.5} L_{\odot}$
only (with a correction for the contribution of star formation
to $L_{O3}$).
\label{fig3}}
\end{figure}

\clearpage

\begin{figure}
\plotone{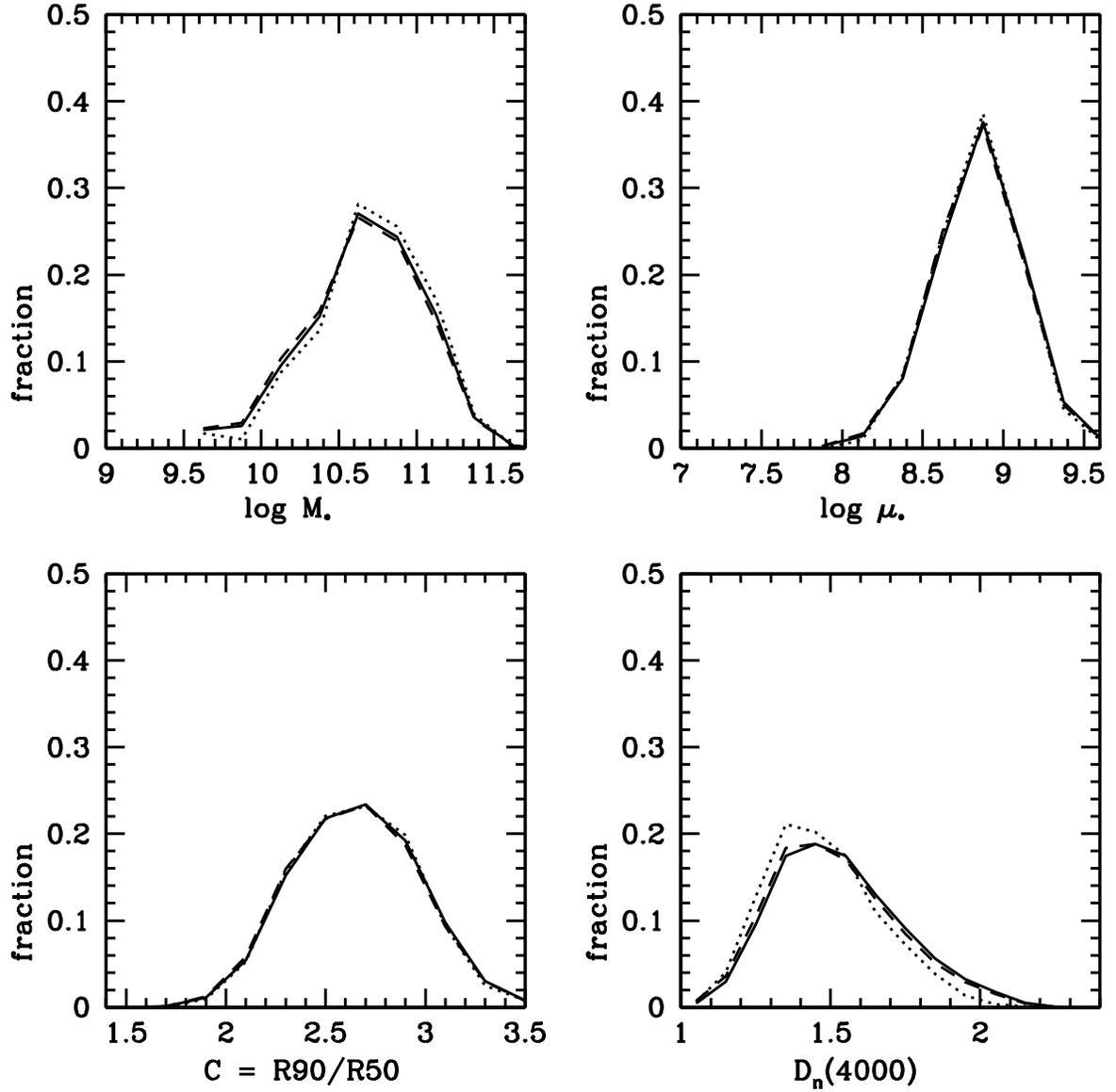}
\caption{The distributions of the total volume-weighted [OIII]
luminosity in Type 2 AGN over $\log M_*$, $\log \mu_*$, $C$, and D$_n$(4000).
The description of the line types is the same as in Fig. 1.
\label{fig4}}
\end{figure}

\clearpage

\begin{figure}
\plotone{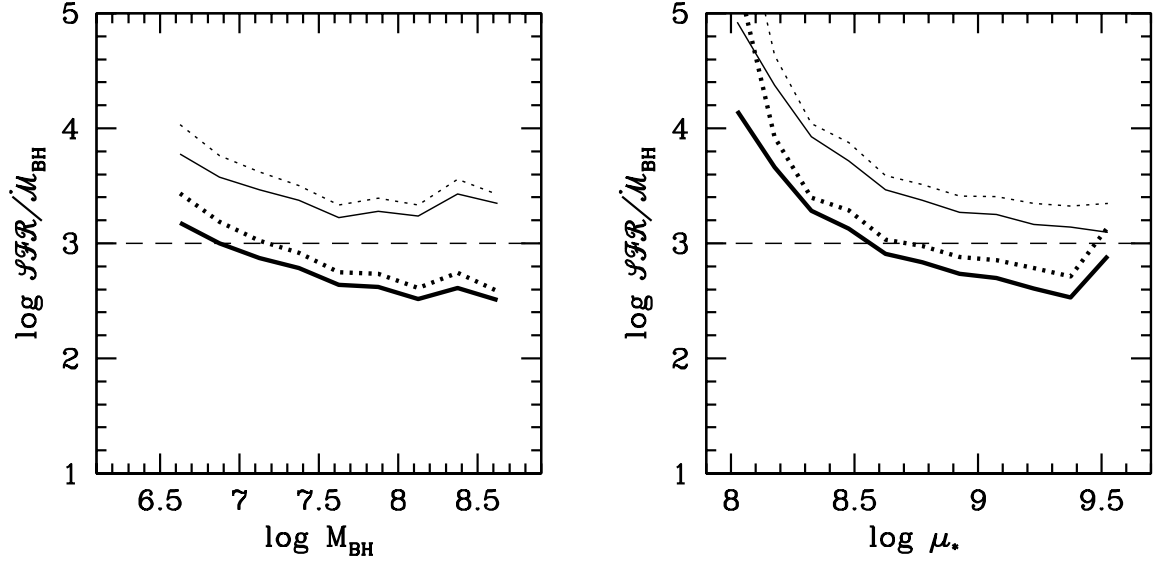}
\caption{The ratio of the total volume-weighted star formation rate in
galaxies to the total volume-weighted accretion rate onto black holes
as traced by Type 2 AGN is plotted as a function of the black hole mass and the galaxy stellar
surface mass density.  The thick black lines show the result if SFR is
calculated within the fiber aperture for each galaxy and the thin
black lines show the result using estimates of {\em total} SFR.  The
dotted lines show results restricted to the AGN with $\log L_{O3} >
10^{6.5} L_{\odot}$.  The dashed line shows the fiducial value of
stellar to black hole mass in bulges and elliptical galaxies.
\label{fig5}}
\end{figure}

\clearpage

\begin{figure}
\plotone{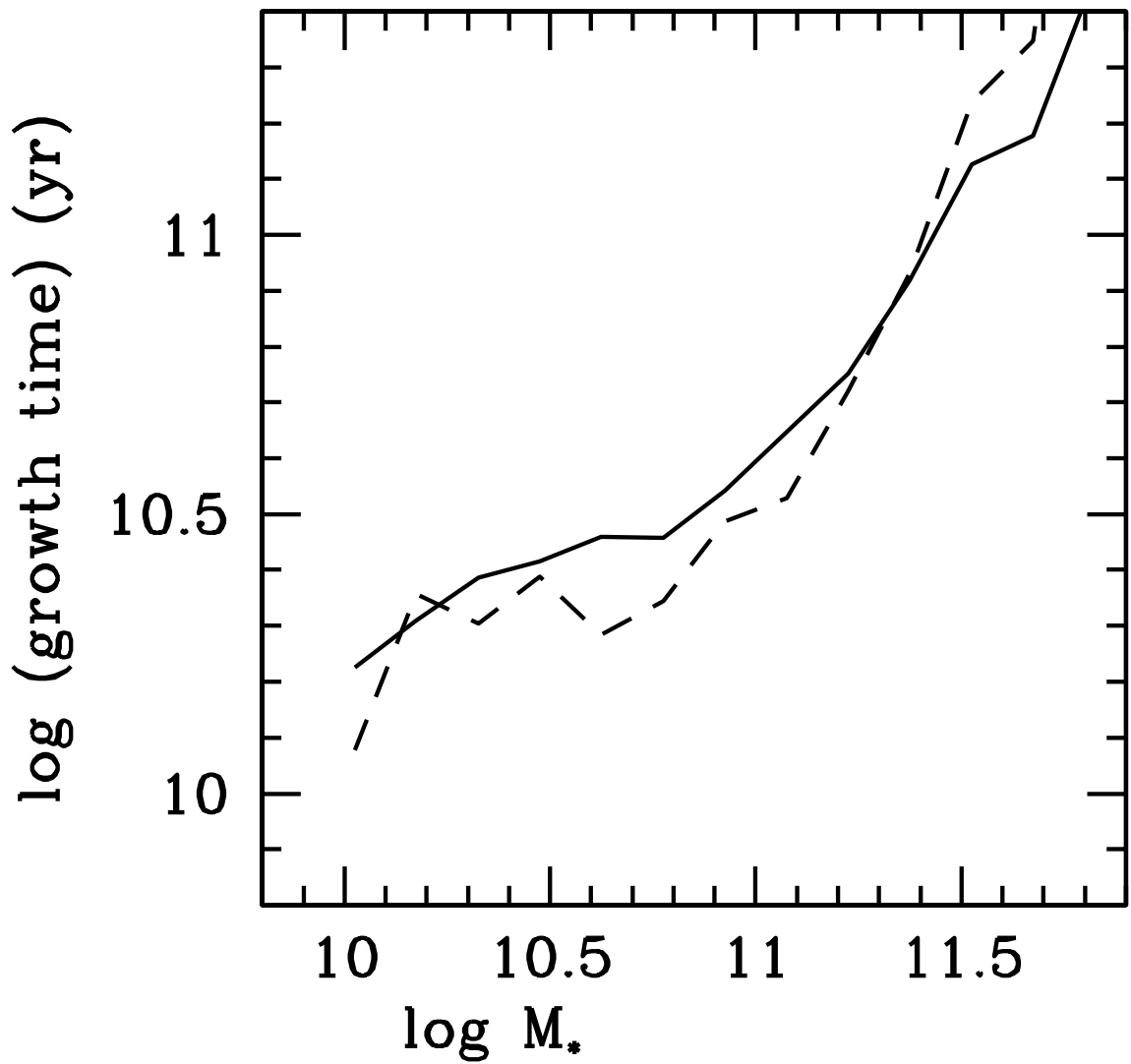}
\caption{The volume-averaged growth time of the galaxy population
calculated within the SDSS fiber aperture (solid) is compared with
that of the black hole population (dashed). Results have been plotted
as a function of galaxy mass. Only AGN with $\log L_{O3} > 10^{6.5} L_{\odot}$
have been included in the estimation of black hole growth.
\label{fig6}}
\end{figure}






\end{document}